\documentclass[aps,prl,reprint,showkeys,superscriptaddress,longbibliography,eprint,articletitle=true]{revtex4-1}
\usepackage[utf8]{inputenc}
\usepackage[english]{babel}
\usepackage{amsmath}
\usepackage{amsfonts}
\usepackage{amssymb}
\usepackage{graphicx}
\usepackage[caption=false]{subfig}
\usepackage{tikz,pgfplots}
\usetikzlibrary{shapes,snakes}
\usepackage{epstopdf}
\usepackage{color}%
\usepackage{enumerate}
\usepackage[hmargin=3cm,vmargin=3cm]{geometry}
\usepackage{marvosym}
\providecommand{\U}[1]{\protect\rule{.1in}{.1in}}
\providecommand{\U}[1]{\protect\rule{.1in}{.1in}}

\usepackage{url,hyperref}
\hypersetup{colorlinks=true,linkcolor=red,
urlcolor=blue!50!black,citecolor=blue!50!black,hypertexnames=false}

\begin{document}
\title{Numerical Simulation and the Universality Class of the KPZ Equation for Curved Substrates}

\author{Roya Ebrahimi Viand}
\affiliation{Department of Physics, Institute for Advanced Studies in Basic Sciences, P.O. Box 45195-1159, Zanjan, Iran}
\affiliation{Freie Universit\"{a}t Berlin, Institute of Mathematics, Arnimallee 6, 14195 Berlin, Germany}

\author{Sina Dortaj}
\affiliation{Department of Physics, Institute for Advanced Studies in Basic Sciences, P.O. Box 45195-1159, Zanjan, Iran}
\affiliation{Freie Universit\"{a}t Berlin, Institute of Mathematics, Arnimallee 6, 14195 Berlin, Germany}

\author{Seyyed Ehsan Nedaaee Oskoee}
\email{nedaaee@iasbs.ac.ir}
\affiliation{Department of Physics, Institute for Advanced Studies in Basic Sciences, P.O. Box 45195-1159, Zanjan, Iran}
\affiliation{Research Center for Basic Sciences \& Modern Technologies (RBST), Institute for Advanced Studies in Basic Sciences (IASBS), Zanjan 45137-66731, Iran}

\author{Khadijeh Nedaiasl}
\affiliation{Department of Mathematics, Institute for Advanced Studies in Basic Sciences, P.O. Box 45195-1159, Zanjan, Iran}

\author{Muhammad Sahimi}
\affiliation{Mork Family Department of Chemical Engineering and Materials Science, University of Southern California, Los Angeles, CA 90089-1211, USA}

\begin{abstract}
	
	The Kardar-Parisi-Zhang (KPZ) equation for surface growth has been analyzed for
	over three decades. The vast majority of the past studies were, however, 
	concerned with surface growth that start from flat substrates. In several 
	natural phenomena and technological processes interface growth occurs on curved
	surfaces. Examples include tumor and bacterial growth, and the interface 
	between two fluid phases during radial displacement of a fluid by another fluid
	in a porous medium. Since in growth on flat substrates the linear size of the 
	system remains constant, whereas it increases in growth on curved substrates, 
	the universality class of the resulting growth process has remained 
	controversial. Some experiments indicated the power law for the interface 
	width, $w(t)\sim t^\beta$, remains the same as in growth on planar surfaces. 
	Escudero (Phys. Rev. Lett. {\bf 100}, 116101, 2008) argued, however, that for the 
	radial KPZ equations in (1+1)-dimension $w(t)$ should increase as $w(t)\sim
	[\ln(t)]^{1/2}$ in the long-time limit. Krug (Phys. Rev. Lett. {\bf 102}, 
	139601, 2009) argued, however, that the dynamics of the interface must remain 
	unchanged with a change in the geometry. Other studies indicated that for 
	radial growth the exponent $\beta$ should remain the same as that of the planar
	case, regardless of whether the growth is linear or nonlinear, but that the 
	saturation regime will not be reached anymore. We present the results of 
	extensive numerical simulations in (1+1)-dimensions of the radial KPZ equation,
	starting from an initial circular substrate. We find that unlike the KPZ 
	equation for flat substrates, the transition from linear to nonlinear 
	universality classes is not sharp. Moreover, in the long-time limit the 
	interface width exhibits logarithmic growth with the time, instead of 
	saturation. We also find that evaporation dominates the growth process when 
	the coefficient of the nonlinear term in the KPZ equation is small, and that 
	the average radius of the interface decreases with time and reaches a minimum 
	but not zero value.
	
\end{abstract}
\maketitle
\newpage

Surface growth is a phenomenon common to many processes of fundamental 
scientific interest and practical applications, and occurs over a broad range 
of length scales, from nanometers in biological growth \cite{bacteria} and fabrication of 
thin films \cite{thinfilm}, to meters or largers in fluid flow in porous media \cite{porousmedia}.
Therefore, the physics of interface growth and the mechanisms that contribute 
to it have been studied for a long time \cite{barabasi,meakin1993} by experiments, theoretical
analysis, and numerical simulations. In particular, interface dynamics has 
been studied with both discrete and continuum models. The former are governed 
by growth rules that are set such that they produce the interface dynamics and 
growth in the phenomena of interest. If the growth phenomenon is to be studied 
at large length scales, the process is modeled by continuum models, represented
by stochastic differential equations \cite{barabasi,meakin1993}. Scaling analysis of such phenomena 
is also very useful, because their dynamics may be characterized by power laws 
and universality classes. In particular, one of the most important properties 
of a growing surface is the interface width $w(t)$, defined by
\begin{equation}
w(t)=\left(\left\langle h^2 \right\rangle - \left\langle h \right\rangle^2 
\right)^{\frac{1}{2}}.
\end{equation} 
where $h({\bf x},t)$ is the height of the surface at position {\bf x} at time 
$t$, and $\left\langle\cdot\right\rangle$ is its average over various 
positions. For most growth phenomena that start from a flat substrate, the 
width $w(t)$ follows the Family-Vicsek scaling law \cite{FamilyVicsek},
\begin{equation}\label{eq:FV}
w(t)\sim \left\{
\begin{aligned}
t^{\beta} \hspace{1cm} t < t_{\times}\\
L^{\alpha} \hspace{1cm} t > t_{\times}
\end{aligned}
\right.
\end{equation} 
in which $\alpha$ and $\beta$ are, respectively, the roughness and growth 
exponents, and $t_{\times}$ is a crossover time, the time at which a transition
occurs from growth to one in which $w(t)$ no longer grows with time. 

One of the most successful continuum models for describing a wide range of 
interface growth is the Kardar-Parisi-Zhang (KPZ) equation \cite{barabasi, kpz}:
\begin{equation}\label{eq:kpz}
\frac{\partial h(x,t)}{\partial t}=\nu\nabla^2h(x,t)+\frac{\lambda}{2}
\left[\nabla h(x,t)\right]^2+F+\eta(x,t),
\end{equation}
in which $F$ is the constant flux of incoming particles (velocity of the height
growth), $\nu$ is the surface tension, $\lambda$ is a parameter of the model, 
and $\eta$ represents thermal fluctuations of the incoming flux with the 
statistical properties,
\begin{eqnarray}\label{eq:eta}
&\langle \eta(x ,t)\rangle = 0, \\
&\langle \eta(x ,t)\eta(x ^{\prime},t^{\prime})\rangle =2D\delta(x -x ^{\prime})\delta(t-t^{\prime}).
\end{eqnarray}
One can choose a reference frame that moves in the growth direction with a 
constant velocity equal to the growth rate of the average height $\langle h
\rangle$, as a result of which Eq. (\ref{eq:kpz}) reduces to
\begin{equation}\label{eq:kpz_red}
\frac{\partial h(x,t)}{\partial t}=\nu\nabla^2h(x,t)+\dfrac{\lambda}{2}
\left[\nabla h(x,t)\right]^2+\eta(x,t)\;.
\end{equation}
Equation (\ref{eq:kpz_red}), the simplest nonlinear model of interface growth, is considered 
the standard model of the phenomenon over the last three decades. Ignoring the 
nonlinear term reduces Eq. (\ref{eq:kpz_red}) to the linear model of Edwards and Wilkinson 
(EW), which in the case of growth on a flat substrate yields, $\beta=(2-d)/4$ 
and $\alpha=(2-d)/2$ for $d-$dimensional growth. Unlike the EW equation, 
however, due to its nonlinearity the exact solution of the KPZ equation is not 
available in any dimension. In (1+1)-growth on a line, the exact values, 
$\alpha=1/2$ and $\beta=1/3$ have been obtained through scaling analysis and 
renormalization group methods \cite{barabasi}, and by random-matrix theory \cite{sasamoto}. In higher 
dimensions, however, the exponents can not be determined analytically and, 
therefore, numerical solutions is used.

In many natural phenomena and technological processes interface growth occurs 
on curved surfaces, such as cylindrical and spherical surfaces. Examples 
include tumor \cite{saberi1,saberi2} and bacterial growth, and the interface between two fluid 
phases when a fluid is injected into a porous medium in which the growth occurs
radially. There are some fundamental differences between such processes and 
growth processes that commence from flat surfaces. A main difference is that in
growth on a flat substrate, the linear size of the system - the end-to-end 
distance between the boundaries - remains constant with time during the 
process, whereas it increases in the case of growth on curved substrates \cite{curved}.

To understand such differences, enlarging substrates have been modeled and 
studied \cite{carasco_kpz,alves2018,alves2013,Carrasco_2014}. A most important issue is whether such differences affect the 
scaling properties of the growth. In order to address the issue, one must 
solve the KPZ equation in cylindrical or spherical geometry with the 
appropriate initial and boundary conditions. Maritan {\it et al.} \cite{maritan}
presented a parametrization-independent form of the KPZ equation, from which 
one obtains the equation for radial growth in (1+1)-dimension \cite{maritan,escudero,kapral,batchelor},

\begin{widetext}
\begin{equation}\label{eq:Radialkpz}
\frac{\partial R(\theta ,t)}{\partial t}=\frac{\nu}{R(\theta ,t)^{2}}\frac{\partial ^{2}R(\theta ,t)}
{\partial \theta ^{2}}-\frac{\nu}{R(\theta ,t)}+F+\frac{F}{2R(\theta ,t)^{2}}\left[
\frac{\partial R(\theta ,t)}{\partial \theta}\right]^2+\frac{1}{\sqrt{R(\theta ,t)}}\eta(\theta ,t),
\end{equation}
\end{widetext}
in which $R(\theta,t)$ is the angle- and time-dependent radius, and 
$\eta(\theta,t)$ is the uncorrelated thermal noise that follows the same 
statistics as in Eqs. (4) and (5), with {\bf x} replaced by $\theta$. 

The dynamics of the radial KPZ equation has been studied in the past. Some 
experiments \cite{universalfluctuations} indicated the same power laws as in the case of growth on 
planar surface, but the issue has remained unresolved. Escudero \cite{escudero}
proposed some analytical conjectures and argued that for the radial EW and KPZ 
equations in (1+1)-dimension, the interface width increases as $w(t)\sim
[\ln(t)]^{1/2}$ in the long-time limit, i.e., when $t\gg t_{\times}$, in 
contrast with the planar case. Krug \cite{krug} argued, however, that the dynamics of 
the interface must remain unchanged with a change in the geometry. Other 
studies \cite{Singha_2005, Khorrami} indicated, however, that for radial growth the exponent 
$\beta$ should be the same as that of the planar case, regardless of whether
the growth is linear or nonlinear, but that the saturation regime will not be 
reached because the mean radius of the interface increases with higher speed 
than the correlation length. These works employed asymptotic analysis in 
combination with simple discrete models, such as the Eden model, in which 
radial growth was introduced by enlarging the lattice with time at the 
interface position. In addition, Batchelor {\it et al.} \cite{batchelor} introduced a form 
of the KPZ equation that exhibited the power law $t^\beta$ of Eq. (\ref{eq:FV}), with the
exponent of the planar growth, while Carrasco {\it et al.} \cite{Carrasco2019} investigated 
numerically the radial EW equation.

It is clear that the nonlinear term of the KPZ equation plays a critical role 
in determining its scaling properties. There are two fixed points (FPs) in the 
renormalization group analysis of the KPZ equation in the Cartesian 
coordinates, with one corresponding to $\lambda=0$, and a second one associated
with $|\lambda|>0$. The former is a repelling FP, whereas the latter is an 
attractive FP, indicating that the growth dynamics belongs to the usual KPZ 
universality class for any nonzero $\lambda$. 

In this Rapid communication we report on the results of extensive numerical
simulation of the radial KPZ equation in (1+1)-dimension, in order to determine
its true behavior in the long-time regime. The interface is grown on a circular
substrate, and we investigate the role of the nonlinear term of the equation 
in order to determine whether, similar to the planar growth, it controls the 
transition between the various universality classes.

Since the KPZ contains several parameters, we first reduce their number by 
introducing dimensionless variables:
\begin{equation}\label{eq:Dless}
\begin{split}
\widetilde{R} &= R \big(\frac{\nu}{D}\big),\\
\widetilde{t} &=t\bigg(\frac{\nu ^{3}}{D^{2}}\bigg),\\
\lambda       &= F\big(\frac{D}{\nu ^{2}}\big).
\end{split}
\end{equation}
Rewriting Eq. (\ref{eq:Radialkpz}) using (\ref{eq:Dless}) and deleting the tilde lead to the following 
equation
\begin{widetext}
\begin{equation} \label{eq:Dlesskpz}
\frac{\partial R(\theta ,t)}{\partial t}=\frac{1}{R(\theta ,t)^{2}}\frac{\partial ^{2}R(\theta ,t)}{\partial 
\theta ^{2}}-\frac{1}{R(\theta ,t)}+\lambda\\
+\frac{\lambda}{2R(\theta ,t)^{2}}\bigg(\frac{\partial R(\theta ,t)}{\partial \theta}\bigg)^{2} +\frac{1}
{\sqrt{R(\theta ,t)}} \eta(\theta ,t)\;,
\end{equation} 
\end{widetext}
which is solved numerically by {\it both} the finite-difference (FD) and 
finite-element (FE) methods. In the FD method we used fully implicit 
discretization for both time and space, which resulted in
 \begin{equation}\label{eq:fddiscrete}
 \begin{gathered}
\frac{R_i^{n+1}-R_i^n}{\Delta t}=
\frac{1}{(R_i^{n+1})^2}\left[\frac{R_{i+1}^{n+1}-2R_i^{n+1}+R_{i-1}^{n+1}}{(\Delta\theta)^2}\right] \\
-\frac{1}{R_i^{n+1}}+\lambda 
+\frac{\lambda}{2(R_i^{n+1})^2}\left[\frac{R_{i+1}^{n+1}-R_{i-1}^{n+1}}{2\Delta\theta}\right]^2 \\
+\frac{\sqrt{2}}{\sqrt{\Delta t}}\frac{\xi_i}{\sqrt{R_i^{n+1}}}\;,
\end{gathered}
\end{equation}
where $n$ and $i$ are, respectively, the time step and the node numbers, and 
$\xi_i$ is a random number attributed to node $i$, generated by the Box-Muller 
algorithm. The number of grid points were 1024, and the time step was 0.2. In 
order to reduce the noise in the results, one hundred realizations were 
generated and the results were averaged over them. Equation (\ref{eq:fddiscrete}) represents a 
system of nonlinear equations, which we solved by the Newton's and bi-conjugate
gradient methods. To check the accuracy of the numerical procedure, we 
recomputed the results for the linear model of growth. We obtained, $\beta= 0.245 $, 
which is in a good agreement with what was reported by Carrasco 
{\it et al.} \cite{Carrasco2019}.

To further check the accuracy of the numerical results obtained by the FD 
method, we disretized the time via the forward Euler's method and used the FE 
method to discretize the angular variable $\theta$. 
The resulting weak form of 
the equation in suitable functionsl space $V$ (in this case, it is the space of square integrable functions with square integrable first derivative) is as follows,
\begin{equation}\label{eq:weakkpz}
\begin{gathered}
\bigg((R^n)^2\frac{(R^{n+1}-R^n)}{\Delta t},T\bigg)= \\
-\bigg(\frac{\partial R^n}{\partial\theta},\frac{\partial T}{\partial\theta}\bigg)+\bigg(\bigg(-R^n+\lambda(R^n)^2+\frac{\lambda}{2}\left(
\frac{\partial R^n}{\partial\theta}\right)^2\\
+\frac{\sqrt{2}}{\sqrt{\Delta t}(R^n)^{3/2}}\xi\bigg),T\bigg),\quad \forall T \in V\;,
\end{gathered}
\end{equation}
where $T$ is the test function, and $(.,.)$ denotes the inner product in $V$. 
To discretize Eq. (\ref{eq:weakkpz}), finite-dimensional subspace $V_n$ of $V$ was 
considered. In the present problem both the test and trial spaces are 
finite-dimensional space, $V_n={\rm span}\{\varphi_i\}_{i=1}^N$, where each
$\varphi_i$ is a piecewise linear polynomial on the quasi-uniform grid in the 
radial direction, and $n$ is the number of basis functions. Thus, $R^n(\theta)$
is approximated by 
\begin{equation}\label{eq:expRn}
R^n(\theta)=\sum_{i=1}^N c_i^n\varphi_i(\theta)\;,
\end{equation}
The time step and the number of angle segmentations were the same as in the FD 
method, namely, 0.2 and 1024, respectively. The resulting linear system of 
equations was solved with the direct LU decomposition method using the LAPACK 
package \cite{lapack}.   

\begin{figure}[ht]
	\centering
	\includegraphics[scale = 0.28, angle=270]{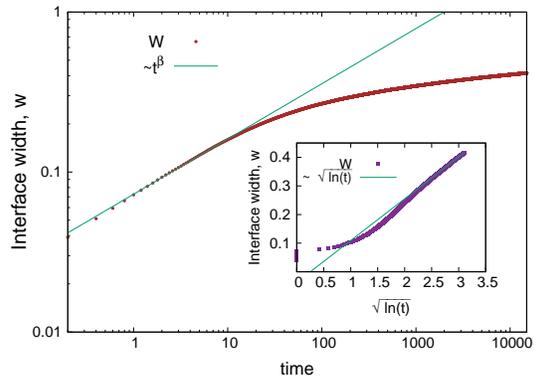}
	\caption{Logarithmic plot of interface width $w$ as a function of time. The 
		fitted curve represents a power law, shown to indicate that the early stages 
		of growth process is indeed governed by a power law. Unlike growth on flat 
		substrates, however, the growth process will continue after saturation, with 
		its speed being very small. This is shown in the inset, which is a plot of $w$ 
		as a function of $\sqrt{\ln t}$.}
		\label{wt1}   
\end{figure}

{\it Results.} Two processes play important roles in the determination of the 
growth dynamics: competition between deposition and evaporation from the 
interface, and saturation, both of which are controlled by the nonlinear term 
of the KPZ equation. Lateral growth, which is a result of the nonlinear term, 
is responsible for the saturation. In growth on a flat substrate, the nonlinear
term induces the appearance of correlations perpendicular to the growth 
direction, which grow with time until the correlation length becomes comparable
with the linear size of the substrate. Subsequently, the entire system becomes 
correlated and, as a result, the roughnening of the surface stops, i.e., $w(t)$
saturates. In contrast, on a spherical substrate, for example, the linear size 
of the system increase with the growing surface's average height, the radius 
of the sphere. Therefore, the correlation length can still grow when the system
saturates, resulting in very slow growth in $w(t)$. This is shown in Fig. \ref{wt1}.
We find, based on the inset of Fig. \ref{wt1}, that in the long-time limit the growth 
is logarithmic, hence supporting the prediction by Escudero and others \cite{escudero,curved}.

\begin{figure}[ht]
	\centering
	\includegraphics[scale = 0.28, angle=270]{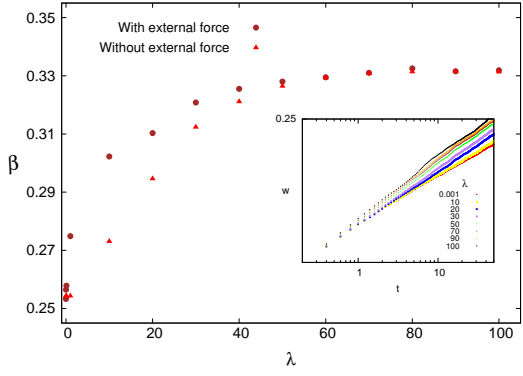}
	\caption{Growth exponent $\beta$ as a function of the coefficient of nonlinear 
		term $\lambda$, computed by numerical simulation of Eq. (\ref{eq:Dlesskpz}). The inset shows
		the exponent $\alpha$.}
	\label{beta}
\end{figure}

Some past numerical simulation of the KPZ equation in (1+1)-dimensions \cite{escudero,universality} 
indicated a universal value, $\beta=1/3$, independent of the numerical value 
of the model's parameters. In the case of growth on a spherical substrate, the 
same value was suggested \cite{escudero, Khorrami}. To compute the numerical value of the growth 
exponent $\beta$, we fitted $w(t)$ to a power law at relatively short times, 
shown in Fig. \ref{wt1}. The results are shown in Fig. \ref{beta}, which presents dependence of 
$\beta$ on the nonlinear coefficient $\lambda$ for two cases, one when the 
constant term in Eq. (\ref{eq:Dlesskpz}), representing an external driving force, takes on 
several values, and the second case when it is zero. In both cases 
$\beta\approx 1/4$ when $\lambda$ is small enough, but not necessarily 0, but
it then approaches 1/3 as $\lambda$ increases. Therefore, unlike the KPZ 
equation in Cartesian geometry, the boundary between the linear and nonlinear 
universality classes is not sharp, rather the system transitions smoothly from 
the linear universality class to the nonlinear one as $\lambda$ increases. When
$\lambda$ is larger than $\approx 80$, however, the system is in the KPZ 
universality class. The transition between these two regimes is characterized, 
however, by a $\lambda-$dependent $\beta$. To make sure that $\beta$ does 
indeed depend of $\lambda$, we also used $w(t)\propto t^\beta f(t)$, where 
$f(t)$ represents the correction terms. We used $f(t)=a+bt^{-\zeta}$, where $a$
and $b$ are constant, and $\zeta$ is a correction-to-scaling exponent, but the
results remained unchanged.

The competition between deposition - adding particles to the interface - and 
evaporation - removing them from the growing surface - is important and is also
controlled by the nonlinear term of the KPZ equation. On a flat substrate, 
positive values of $\lambda$ represent the case in which the number of newly 
deposited particles on the interface is on average larger than those leaving 
it, hence increasing the average height $\langle h\rangle$, even in the absence
of the external driven force. In contrast, $\lambda<0$ implies evaporation 
that takes out more particles than those added by deposition, hence 
shrinking $\langle h\rangle$.

\begin{figure}[ht]
	\centering
	\includegraphics[scale = 0.28, angle=270]{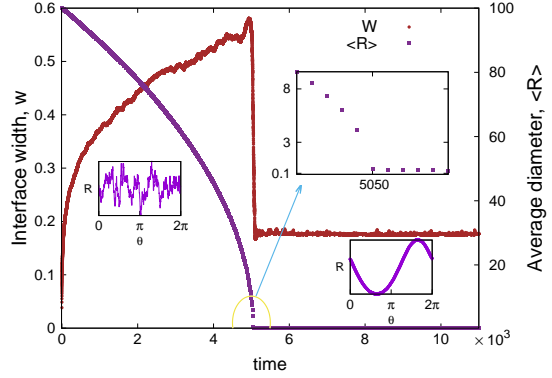}
	\caption{Plot of interface width $w$, as well as average interface diameter 
		$\langle R\rangle$, versus time for $\lambda=10^{-3}$. Top right inset: a 
		zoomed-in part of $\langle R\rangle$ at the transition point. The other two 
		inset plots are snapshots of the interface at times before (left) and after 
		(right) the transition point.}
	\label{RW}
\end{figure}

But in spherical growth $\lambda>0$ does not necessarily increase $\langle h
\rangle$. This is shown in Fig. \ref{RW} that presents plots of both the average 
radii of the interface $\langle R(t)\rangle$ and the interface width $w$ for 
$\lambda=10^{-3}$. Here, $\langle R\rangle$ is equivalent to $\langle h\rangle$
in growth on flat substrate. It is clear that surface thickness decreases with 
time until it reaches a nonzero minimum value. The top right inset in Fig. \ref{RW} is
a zoomed-in view of $\langle R\rangle $ at the transition point. The interface 
width has its usual power-law growth before the transition point, but it 
suddenly drops to a relatively small value at the transition point and, then, 
fluctuates around it. 

To understand the transition better, we plotted two snapshots of the interface,
one at a time earlier than the transition - the inset on the left side of Fig. 
\ref{RW} - and the other one after the transition - the right bottom inset. Before the
transition, the interface has a usual roughness profile, as reported in many of
numerical simulation of the KPZ equation. The morphology of the interface is, 
however, very smooth after the transition. A remarkable aspect of the growth 
process at this point is that both $\langle R(t)\rangle$ and $w(t)$ reach a 
fixed point and fluctuate around it. Even thermal noise does not play any role 
in this case, as it cannot make any significant change in either the morphology
of interface or in its average height.

\begin{figure}[ht]
	\centering
	\subfloat{\includegraphics[scale = 0.28, angle=270]{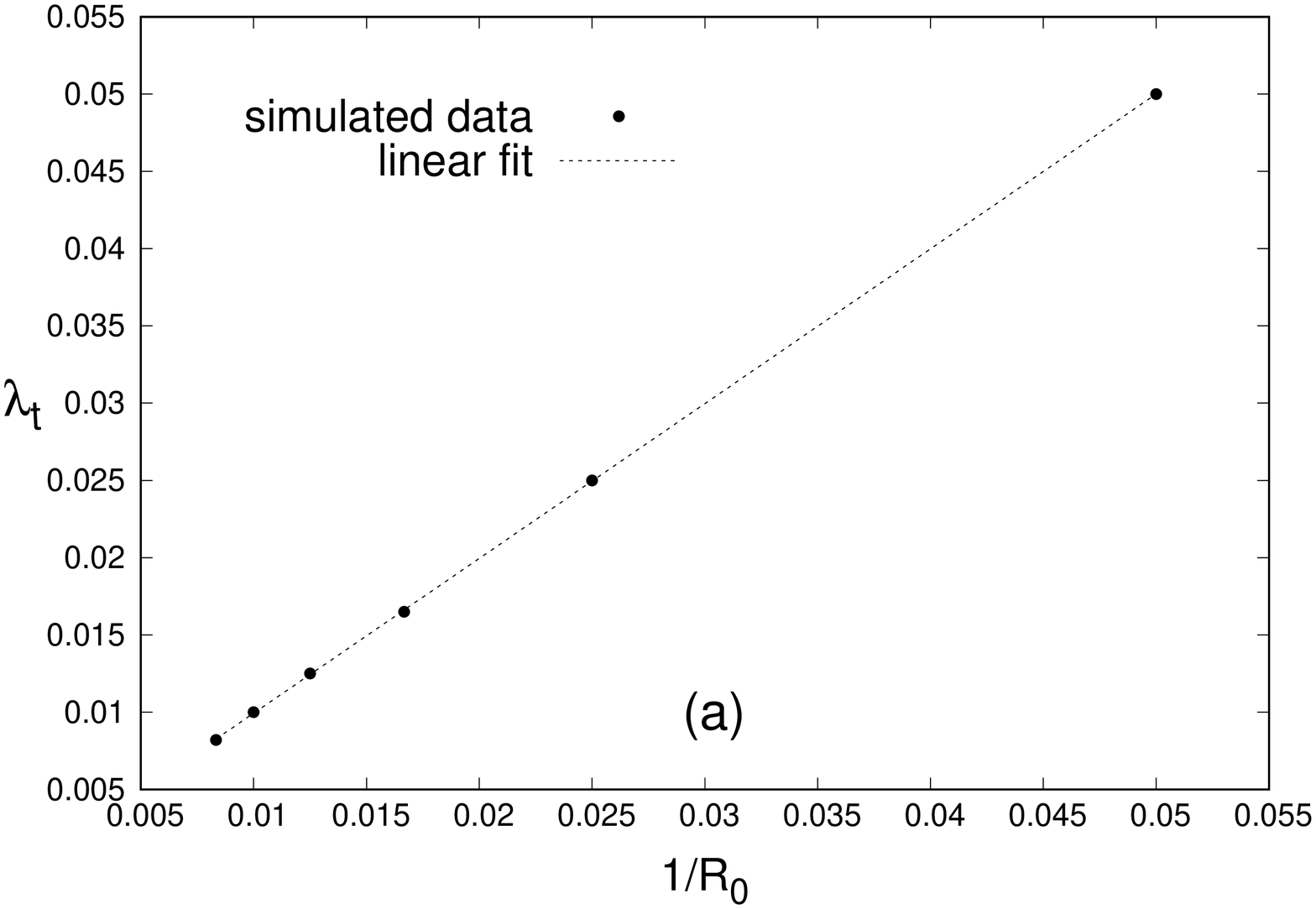}}\qquad
	\subfloat{\includegraphics[scale = 0.28, angle=270]{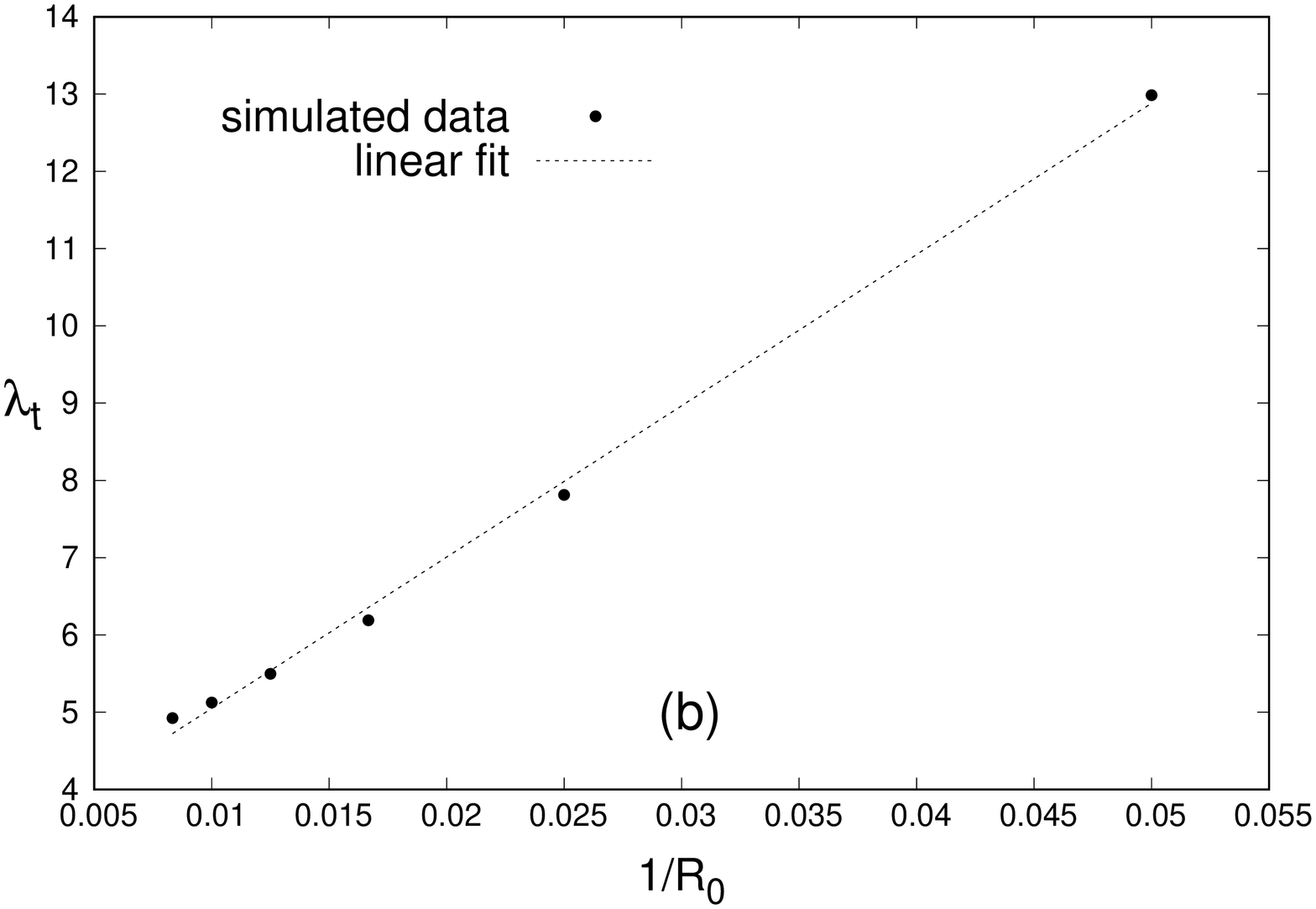}}
	
	\caption[]{The transition value of the $\lambda$ as a function of inverse of 
		the initial radii $R_0$ of substrate for, (a) with a nonzero constant term, and
		(b) when it is zero.}
	\label{landatransition}
\end{figure}

As mentioned earlier, the transition between evaporation- and 
deposition-dominated regimes occurs at $\lambda=\lambda_t$, which depends on 
the curvature of the substrate. Since we start the growth on a substrate 
without fluctuations, all terms Of Eq. (\ref{eq:Dlesskpz}) containing derivatives do not 
contribute at the initial stage of the growth, and the governing equation 
reduces to the following form:
\begin{equation}\label{eq:kpznoderivative}
\frac{\partial R(t,\theta)}{\partial t}=-\frac{1}{R(t,\theta)}+\lambda+
\frac{1}{\sqrt{R(t,\theta)}}\;\eta(t,\theta)\;.
\end{equation}
Since the average of the thermal noise is zero, the right side of Eq. (\ref{eq:kpznoderivative}) is
positive ($R$ increases) when $\lambda=\lambda_t>1/R_0$, with $R_0$ being the
initial radius of the surface when the growth begins. This is shown in Fig. \ref{landatransition},
which presents $\lambda_t$ as a function of $1/R_0$ for two cases, one without 
the constant term, and the second one with a nonzero term. They both appear to 
be linear functions of $1/R_0$ and, therefore, were used to obtain the 
asymptotic value for a flat substrate, the limit $R_0\to\infty$. We obtained, 
$\lambda_t\to 0$ as $R_0\to\infty$, indicating that in that limit Eq. (\ref{eq:Dlesskpz}) 
becomes equivalent to the standard KPZ equation. In contrast, when the constant
term is dropped out, the asymptotic value is $\lambda_t\approx 3$, which 
differs from that of the standard KPZ equation. 

To further explain the difference between the two cases, we recall the role of 
the constant term in the standard KPZ equation. Ignoring the term for growth on
a flat substrate implies that an observer is in the reference frame, moving 
with a constant velocity, but viewing it in a reference frame that is fixed on 
the substrate, resulting in the same observations. In the case of the growth 
on a curved substrate, however, ignoring the constant term does not imply a 
reference frame moving with the average growth velocity, hence it provides a 
different vision for the interface dynamics. 

\begin{figure}[ht]
	\centering
	\includegraphics[scale = 0.28, angle=270]{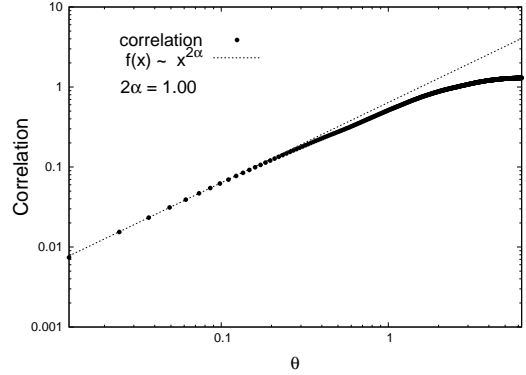}
	\caption{The correlation function $C$ versus $\theta$ for $\lambda=10$ at the 
		final stage of the simulation. Dashed line represents the fitted power law 
		with an exponent $2\alpha$.}
	\label{corr}   
\end{figure}

To determine the universality class of the dynamics of Eq. (\ref{eq:Dlesskpz}), one must 
compute both the growth exponent $\beta$ and the roughness exponent $\alpha$. 
The latter was computed using the height-height correlation function:
\begin{equation}
C(|\theta_1-\theta_2|)=\left\langle|R(\theta_1)-R(\theta_2)|^2\right\rangle\;,
\end{equation}
where $\langle\cdot\rangle$ indicates on average over the thermal noise. 
$C(\theta)$ has the following power-law form,
\begin{equation}
C(\theta)\sim\theta^{2\alpha}
\end{equation}
where $\alpha=1/2$ for the standard KPZ equation \cite{barabasi,kpz,halpin}. Figure \ref{corr} is a 
typical plot of correlation function at the final stage of simulation.

The results are listed in Table 1 for both cases of with or without the 
constant term. In both cases, when $\lambda>\lambda_t$, $\alpha$ is compatible 
with that of the standard KPZ equation. On the other hands, $\alpha\approx 1$ 
when $\lambda<\lambda_t$, indicates a completely correlated and smooth 
interface, as shown in the right bottom inset of Fig. \ref{RW}. When considering the 
constant term, $C(\theta)$ reaches its flat regime at very small $\theta$, 
leaving a very small set of data for fitting, hence estimates of $\alpha$ 
unreliable for $\lambda>1$.

\begin{table}[ht]
	\centering
	\begin{tabular}{|c|c|c|}
		\multicolumn{1}{c}{} & \multicolumn{1}{c}{with constant term} & \multicolumn{1}
		{c}{without constant term} \\ 
		\hline
		$\lambda$ & $2 \alpha$ & $2 \alpha$ \\
		\hline
		0.001 & 1.99 & 1.99\\			
		0.01  & 1.02 & 1.99\\			
		0.1   & 0.95 & 1.99\\			
		1     & 0.87 & 1.99\\							
		3     & \Cutline & 1.99	\\			
		5     & \Cutline & 1.10	\\			
		7     & \Cutline & 0.98	\\			
		9     & \Cutline & 1.10	\\			
		10    & \Cutline & 1.00	\\
		\hline
	\end{tabular}
	\caption{ \label{tab1}}
\end{table}

{\it Summary.} The vast majority of the previous studies of the KPZ equation 
for surface growth over the past three decades were concerned with surface 
growth that begin from flat substrates. In several natural phenomena and 
technological processes interface growth occurs on curved surfaces. Since in 
growth on flat substrates the linear size of the system remains constant, 
whereas it increases in the case of growth on curved substrates, the 
universality class of the resulting growth process has remained controversial, 
with conflicting results reported by various groups. We presented the results 
of extensive numerical simulation in (1+1)-dimensions of the KPZ equation, 
starting from an initial circular substrate. Our results indicate that, unlike 
the KPZ equation for flat substrates, the transition from linear to nonlinear 
universality classes is not sharp. Moreover, the interface width exhibits 
logarithmic growth with the time, instead of saturation, in the long-time 
limit. In addition, the simulations indicate that evaporation dominates the 
growth process when the coefficient of the nonlinear term in the KPZ equation 
is small, and that average radius of interface decreases with time and reaches 
a minimum but not zero value. Thus, while there are certain similarities 
between surface growth on flat and curved surfaces, there are also significant 
differences.


%

\end{document}